\documentclass[12pt,a4paper]{article}

\usepackage{amsmath}
\usepackage{graphicx}
\usepackage{times}
\usepackage{cite}
\usepackage{ulem}
\begin{document}
\begin{titlepage}

\title{Central   elastic  scattering}
\author{ S.M. Troshin\footnote{Sergey.Troshin@ihep.ru, +79151953408}, N.E. Tyurin\\[1ex]
\small  \it NRC ``Kurchatov Institute''--IHEP\\
\small  \it Protvino, 142281, Russian Federation,\\
\small This article is registered under preprint number /hep-ph/2101.07504
%\small Sergey.Troshin@ihep.ru
}
\normalsize
\date{}
\maketitle

\begin{abstract}
   We comment on  phase selection of the scattering amplitude,  emphasizing that the elastic overlap function should have a central impact parameter profile at high energies  and highlighting the role of the reflective scattering  mode at the LHC energies. 
    Emerging problems with the use of peripheral impact parameter dependence of the elastic overlap  function are explicitly indicated. Their solution is an elimination of the  phases connected to  peripheral form of the elastic overlap function. Contrary, we adhere to a relative peripheral form of the {\it inelastic} overlap function with an additional new feature of a maximum at nonzero value of the impact parameter at the highest energies.  Phenomenologically, the dynamics of hadron scattering is motivated by a hadron structure with  a hard central core presence. 
\end{abstract}
Keywords: scattering amplitude, phase, unitarity, analyticity, 
arXiv: 2101.07504.
\end{titlepage}
\setcounter{page}{2}
\section*{Introduction. Role  of the scattering amplitude phase}
Phase of the elastic scattering amplitude plays an important role in physics interpretation of hadron scattering. It is just as important as  the scattering amplitude modulus. Unfortunately,   the phase is an essentially unknown quantity, and the experimental data on the differential cross--section of the elastic scattering allow one to reconstruct  the amplitude $F(s,t)$ with its  phase at $-t=0$ only with the use of   Coulomb--Nuclear Interference (CNI) contribution at very small values of $-t$. Moreover, certain model assumptions are needed  even at  this reconstruction. In the simplest case of assumed constant (i.e. $t$--independent) phase the relation of the real to imaginary parts  ratio of the forward scattering amplitude $\rho$ with the phase $\arg F(s,t)$ is the following:
\[
\arg F(s,t)=\frac{\pi}{2}-\arctan \rho.
\]
However, the phase $t$--dependence can be a nontrivial one and therefore
several various parameterizations of the nuclear amplitude phase dependencies on $t$ have been used in \cite{totem}. A significant uncertainty allows one to use a wide range of assumptions on the phase behavior and 
 it would be useful to find some  arguments limiting  the phase selection.
 
 A role of the phase in hadron scattering is under active theoretical discussion nowadays, cf. e.g. \cite{dre,petr,khoze,dur,ferr,las}, in particular, due to its relation to the recent measurements of the real to imaginary part ratio  $\rho$.

 The elastic and inelastic overlap functions $h_{l, el}$ and $h_{l,inel}$ introduced by Van Hove \cite{vanh} are related by unitarity and are strongly interdependent with the phase. The  important role of the phase has been clearly demonstrated under analysis of the LHC experimental data obtained  by the TOTEM Collaboration \cite{totem}. 

 To this end, it is instrumental to address the elastic and inelastic overlap functions in order to limit choice of the phases  suitable for further considerations and phenomenological analysis. It is aim of  present note.

Thus, we discuss an exclusion  of the phase $t$-dependencies  on the ground of unitarity and analyticity of the scattering amplitude. Peripheral distribution of elastic scattering has been suggested by  analogue tunneling  picture of strong interactions based on the hypothesis of "maximal importance" of particle production \cite{fsch}. The  cases, subject to elimination, correspond to a peripheral form of the elastic overlap function  \cite{krupa}. 
Despite  the explicit form of the phase dependence after elimination of peripheral distribution of the elastic scattering remains to be unknown and leave us with results of phenomenological models only, it allows one to reduce  an arbitrariness
under the phase selection in the  data analysis \cite{totem}.

The reasons for  elimination of the peripheral option of $h_{l,el}$ at phase preselection stage  are discussed in Section 1. Section 2 is devoted to the central profile of the elastic overlap function in the impact parameter representation and its correlation with the reflective scattering mode.  The results come from a combined utilization of unitarity and analyticity in the region of large impact parameters.

\section{Problems with  a peripheral form of the elastic overlap function}

The  partial elastic scattering matrix element can always be represented  as  complex function:
\begin{equation}\label{sm}
S_l(s)=\kappa_l (s)\exp[2i\delta_l(s)]
\end{equation}
with the two real functions $\kappa_l$, $\delta_l $.  Herewith $\kappa_l$ can vary in the interval
$0\leq \kappa_l \leq 1$ and is known as an absorption factor.  The value $\kappa_l =0$ means a complete absorption of the corresponding initial state.

The function $h_{l, el}(s)\equiv |f_l(s)|^2$ is a contribution of the elastic intermidiate states while $h_{l, inel}(s)$ is  a contribution of the inelastic intermediate states into the unitarity relation: 
\begin{equation} \label{unfl}
\mbox{Im} f _l(s)=h_{l,el}(s)+ h_{l,inel}(s).
\end{equation}
The latter can be expressed through the function $\kappa_l(s)$ by  the relation
\begin{equation}
\kappa_l^2 (s)=1-4h_{l,inel}(s).
\end{equation}
The normalization is such that $S_l=1+2if_l$.
In Eq. (\ref{unfl})  $f_l(s)$ is the partial scattering amplitude,  $h_{l, el}(s)$ and  $h_{l, inel}(s)$   correspond to the elastic and inelastic overlap functions $h_{el}(s,b)$ and  $h_{inel}(s,b)$   which can be related to probability distributions of the elastic and inelastic interactions over the impact parameter $b=2l/\sqrt{s}$ (cf. \cite{webb}).

 Evidently, the unitarity relation  implies that the limiting behavior $\mbox{Im}f_l\to 1$ leads to a vanishing real part of the scattering amplitude, i.e. $\mbox{Re}f_l\to 0$,  cf. \cite {tr}.

 Eq. (\ref{unfl})  can be rewritten in the form
\begin{equation} \label{unfl1}
\mbox{Im} f _l(s)[1-\mbox{Im} f _l(s)]=[\mbox{Re} f _l(s)]^2+ h_{l,inel}(s).
\end{equation}

We consider the region of $l\gg 1$,  $s\gg 4m^2_\pi$ and large $2l/\sqrt{s}$, i.e. the region of peripheral interactions at high energies. Eq. (\ref{unfl1}) can be simplified in this  region since
$\mbox{Im} f _l(s)\ll 1$. The  smallness of $\mbox{Im} f _l(s)$ results from the axiomatic field theory \cite{mt} (short--range nature of strong interactions), and the unitarity relation can be approximated by the following one:
\begin{equation} \label{unfl2}
\mbox{Im} f _l(s) \simeq [\mbox{Re} f _l(s)]^2+ h_{l,inel}(s).
\end{equation}
An evident upper bound for the real part squared 
\begin{equation} \label{unfl3}
[\mbox{Re} f _l(s)]^2\leq \mbox{Im} f _l(s)
\end{equation}
 assumes that
 any phenomenological model with a nonvanishing real part of the amplitude should  test its  amplitude against this inequality along with the other constraints.

It will be shown  further that unitarity being combined with  Mandelstam analyticity allows one to obtain a more stringent constraint for the profile of the elastic overlap function and, as a consequence, for the scattering amplitude phase. 

Namely, it is known (cf. \cite{pdbc}) that   the amplitude $f_l(s)$ (its real and imaginary parts) decreases exponentially with $l$ at large values of $l$ and $s$ according to the  Froissart--Gribov formula  \cite{frs,grb}:
\begin{equation} \label{unfl4}
f_l(s)\simeq \omega (s)\exp(-\mu\frac{2l}{\sqrt{s}}),
\end{equation}
where $\omega(s)$ is a complex function of energy and $\mu$ is determined by the position of the lowest singularity in the $t$--channel. Thus,   the exponent is the same for the  real as well as for the imaginary parts of the amplitude $f_l$ and  is determined by the mass of two pions.
Eq. (\ref{unfl4}) originates from Mandelstam representation and is consistent also with  analyticity of the  amplitude $F(s,t)$ in the Lehmann--Martin ellipse \cite{mt}.

 We address the problems related to  a peripheral option for the elastic scattering in what follows.
Peripheral dominant role of $h_{l,el}(s)$ has been discussed in \cite{cn,kt,kt1,krupa} and
could occur due to a nontrivial contribution of the real part  at large values of $l$, i.e. by a relevant choice of the scattering amplitude phase. 

If the elastic scattering is assumed to be dominant in a peripheral region and the contribution of the inelastic states can be neglected  in Eq. (\ref{unfl2}), the following approximate  relation should be considered as valid at large values of $l$,$s$ and  $2l/\sqrt{s}$:
\begin{equation} \label{unflp}
\mbox{Im} f _l(s) \simeq [\mbox{Re} f _l(s)]^2.
\end{equation}
Both Eq. (\ref{unflp}) or the elastic unitarity 
\begin{equation} \label{elun}
\mbox{Im} f _l(s) =  h_{l,el}(s)
\end{equation}
 are in conflict with  Eq. (\ref{unfl4}) since the latter points to the same exponent of the decrease with $l$ for the real and imaginary parts of the scattering amplitude. Therefore,  peripheral domination of the elastic scattering should be discarded
with the corresponding elimination of the respective  phases.

Alternative option anticipates a central  elastic scattering. 
Then Eq. (\ref{unfl2}) can be approximated in the following form 
\begin{equation} \label{unflc}
\mbox{Im} f _l(s) \simeq  h_{l,inel}(s),
\end{equation}
confirming a shadow nature of the elastic scattering at large values of $l$. 
The use of the phases, leading to peripheral dominance of the elastic scattering  \cite{totem} looks like an artificial version because of the shadow nature of the  amplitude at large impact parameters takes place (cf. e.g. \cite{pkan,ll}).
Moreover, we can conclude on the following behaviour  of $h_{l,inel}(s)$    
 \begin{equation} \label{hinel}
h_{l,inel}(s) \simeq \mbox{Im}\,\omega (s)\exp(-\mu\frac{2l}{\sqrt{s}})
\end{equation}
in the region of large $l$ and $s$ variation.
In addition, the  relations
\begin{equation} \label{unflp1}
[\mbox{Re} f _l(s)]^2\ll h_{l,inel}(s) 
\end{equation}
and, consequently,
\begin{equation} \label{unflp2}
[\mbox{Re} f _l(s)]^2\ll \mbox{Im} f _l(s)
\end{equation}
are being valid for the large values of $l$, $s$ and $2l/\sqrt{s}$. 

Those relations reflect a shadow nature of the elastic scattering, cf. Eq. (\ref{unflc}),  i.e.  they correspond to the central profile of the elastic scattering in the impact parameter representation. The large--$l$ tail of $\mbox{Im} f _l(s)$ at high energies is due to the inelastic collisions. 

One should conclude that the assumption on approximate elastic unitarity applicability in this region of $l$ and $s$ variation contradicts to the analytical properties of the scattering amplitude. Indeed, this  is  not surprising from the general principles and from viewpoint  based on  the semiclassical scattering picture of hadrons.

\section{Central elastic scattering}
We proceed with  discussion of  the physical picture which corresponds to the central elastic scattering.
It is a common practice to analyze the hadron scattering in the impact parameter representation 
(cf. \cite{halz}) which is a convenient way to use a semiclassical picture of hadron collisions  at high energies \cite{webb, bl,gold}. Indeed, 
the impact parameter $b$  is a conserved quantity at high energies and   the scattering amplitude   is determined by the Fourier--Bessel transformation of the amplitude $F(s,t)$. The elastic and inelastic overlap functions can be interpreted as the differential contributions to the elastic and inelastic cross--sections respectively over the impact parameter $b$ \cite{webb}. 

Large--$b$ behavior of the scattering amplitude can be obtained from the  relation similar to the Froissart--Gribov projection formula \cite{bl,hz,schr}. Note, that due to  high  collision energy  and the exponential decrease of the amplitude $F(s,t)$ (both of the real and imaginary parts) with $-t$, the effect of finite integration range over $-t$    is not significant and has therefore been neglected by the use of the infinite integration limit. 

In what follows a central profile  of the elastic overlap function will be considered.  
At the LHC energies, the respective estimates lead to the value of the elastic overlap function equal to $0.31$ at $b=0$  \cite{totem}\footnote{From another point of view, this value can be used for the estimation of the value of $[\mbox{Re} f(s,b=0)]^2$. It is approximately equal to $0.03$ since $\mbox{Im} f(s,b=0)\simeq 0.53$ \cite{evg} at the energy $\sqrt{s}=7$ TeV. } indicating   transition to the reflective scattering mode \cite{ijmpa07}
 in agreement with the results  of \cite{alkin1,tsrg}. It has been shown that  the inelastic overlap function $h_{inel}(s,b)$   is very close to its limiting value  
$h^{max}_{inel}=1/4$ in the  region of small and moderate impact parameters, $0\leq b\simeq 0.4$ fm at the LHC energy $\sqrt{s}=13$ TeV. Deviation of $h_{inel}$ from its maximal value  is small and negative in this region of the impact parameters and 
\[
h_{el}(s,b)>1/4>h_{inel}(s,b).
\]
In fact, $h_{inel}(s,b)$ has a shallow local minimum at $b=0$.  
Asymptotically in the reflective scattering mode
$h_{el}(s,b)\to 1$
and $h_{inel}(s,b)\to 0$
at $s\to\infty$ and fixed  $b$ values in the central region\footnote{Both limiting values are equal to $1/4$ in case of absorptive scattering mode, i.e. the black disc saturation.}. 

At the LHC energy $\sqrt{s}=13$ TeV, the unitarity relation  at the impact parameters  range $0\leq b\simeq 0.4$ fm  gives 
\begin{equation} \label{bdh}
(\mbox{Im}f(s,b)-{1}/{2})^2 +(\mbox{Re} f(s,b))^2\simeq 0.
\end{equation}
The estimations $\mbox{Re} f(s,b)\simeq 0$ and $\mbox{Im} f(s,b)\simeq 1/2$ result from Eq. (\ref{bdh}). Thus, the impact parameter picture of the elastic scattering combined with the unitarity  can provide at least a qualitative explanation of the recent  result on the unexpectedly small  real to imaginary parts ratio of the forward  scattering amplitude \cite{ro}.   
 
 In fact, there is no room for a significant  real part of the elastic scattering amplitude at $\sqrt{s}=13$ TeV even at higher values of the impact parameters.
The real part of the  amplitude $f(s,b)$ can be   neglected due to its smallness (cf. for the numerical estimations \cite{drem}, the arguments in favor of such an approximation have  been given in Section 1). 

So, the  replacement $f\to i f$ is used
and the function $S(s,b)$ is taken to be real. 
The   $S(s,b)$ can acquire  the negative values at  $b<r(s)$ at high enough energy\cite{ijmpa07}. The $r(s)$ increases as $\ln s$ at $s\to\infty$ and its value at $\sqrt{s}=13$ TeV is approximatery equal to $0.4$ fm. At $b<r(s)$ the reflection appears, i.e the function $S(s,b)$ in Eq. (\ref{sm}) crosses zero at $b=r(s)$ and  the value of the phase $\delta$ changes abruptly from $0$ to 
 $\delta = \pi/2$. 
 
Under the reflective scattering regime, $f>1/2$, an {\it increase} of the elastic scattering amplitude $f$ correlates with {\it decrease} of $h_{inel}$ at small values of $b$ due to the unitarity relation
\begin{equation}
[f(s,b)-{1}/{2}]^2={1}/{4}-h_{inel}(s,b)=\kappa^2(s,b) /4.
\end{equation}
The negative  $S(s,b)$ correspond to  $\delta=\pi/2$\footnote{Despite that experimental data are in favor of the reflective scattering, aka hollowness \cite{las}, the results of the experimental data analysis of the inelastic overlap function are affected by the assumptions on the real part or phase of the scattering amplitude \cite{las,sam}}.  The term ``reflective''  comes from optics
where the phases of incoming and outgoing  waves under reflection differ by $\pi$. Hollowness or $h_{inel}(s,b)$ getting maximum  at $b>0$ results from reflective scattering and vice versa due to the unitarity. Possibility of such impact parameter dependence of the inelastic overlap function was briefly mentioned in \cite{khr} and discussed in \cite{plb,str,hdre,anis,arr}.  Those phenomena are already relevant for the LHC energies.

The  physical picture of  the hadron interaction region  in the transverse plane  at high energies corresponds to
 a reflective
disc surrounded by a   black ring. It should be emphasized that it is a picture of the interaction region and not the one of the individual protons.
Appearance of the reflective ability can be associated with a soft deconfinement proposed in \cite{wf}. 
The reflective scattering can also be associated  with formation of the color conducting medium in the intermediate state of hadron interaction \cite{jpg19}. 

This picture finds its confirmation in the experiments at JLab and at the LHC  \cite{totem,burk}.  
 The results are in favor of the dominance of the elastic scattering in soft deconfined state.
The mechanism is energy--dependent and leads to the  elastic scattering dominance at $s \to \infty$ as a consequnce of its increasing decoupling  from the inelastic production \cite{del}. 

A central profile of the elastic overlap function is encoded into the  relation
\begin{equation}\label{cent}
\frac {\partial h_{el}}{\partial b}=\left( \frac{1-S}{S}\right)\frac{\partial h_{inel}}{\partial b}.
\end{equation}
resulting from \cite{prr}. Note, that  $( {1-S})/{S}<0$ in the reflective scattering region. The respective
qualitative dependencies are presented in Fig. 1.
\begin{figure}[hbt]
	\vspace{-0.42cm}
	\hspace{-1cm}
	%	\begin{center}
	\resizebox{15cm}{!}{\includegraphics{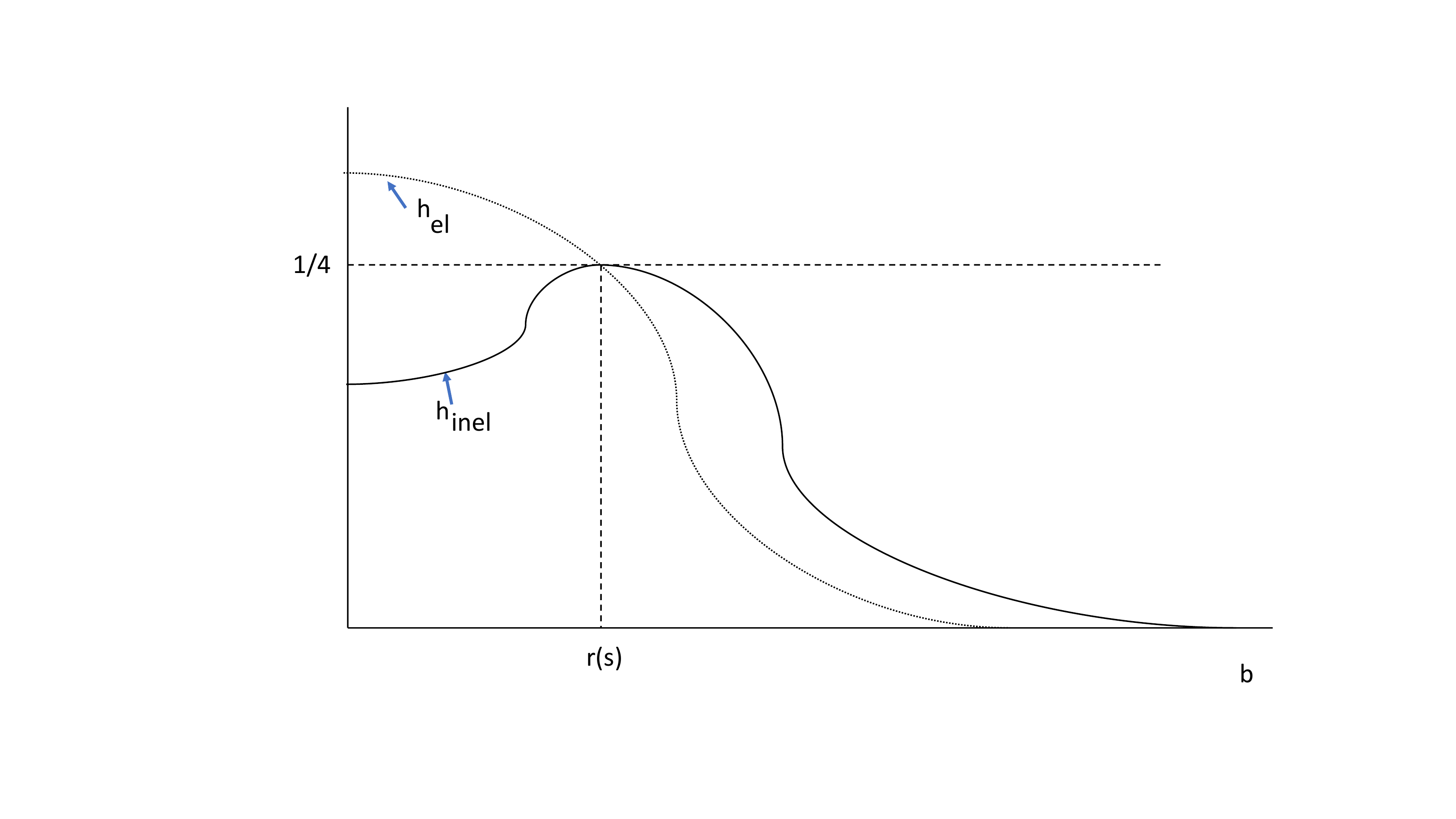}}		
	%\includegraphics{dsdb2n.eps}
	%	\end{center}
	\vspace{-1cm}
	\caption{}	
\end{figure}	 
The picture of hadron scattering is schematically represented at Fig. 2  \cite{epl}. 
  \begin{figure}[hbt]
 	\vspace{-0.2cm}
 	\hspace{-1cm}
 	%	\begin{center}
 	\resizebox{16cm}{!}{\includegraphics{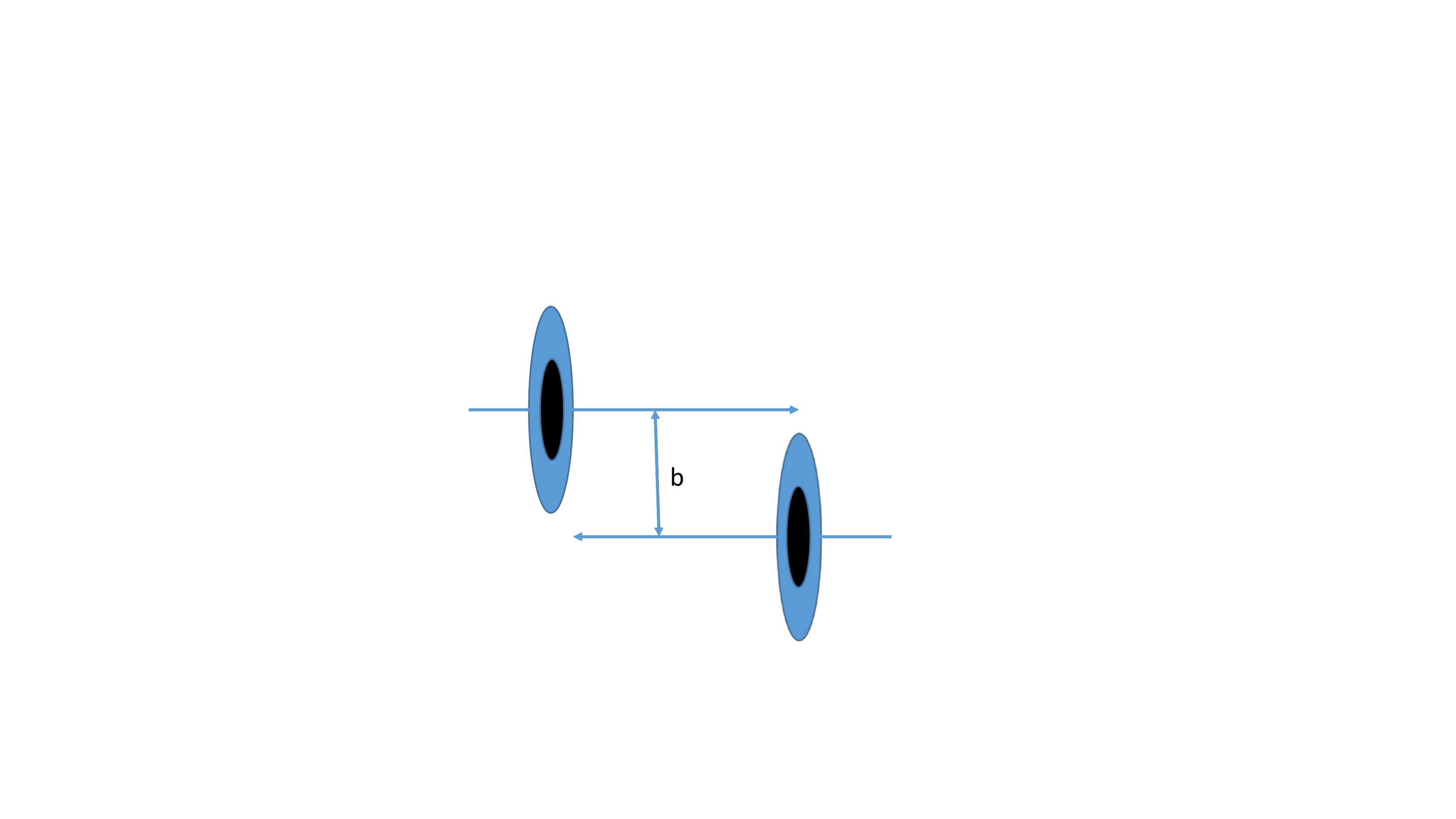}}		
 	%\includegraphics{dsdb2n.eps}
 	%	\end{center}
 	\vspace{-2cm}
\caption{}	
 \end{figure}	 
  Because  of the hard core presence, the elastic scattering of hadrons has a central profile  and looks similar to  collisions of billiard balls. Under the central collisions the cores survive in soft processes.  An inelasticity is  therefore depleted in the central collisions due to unitarity.

 \section{Concluding remarks}
  
 The peripheral profile of the elastic overlap function, i.e. its domination over the inelastic one at large values of $b$, corresponds to assumption of the {\it elastic} unitarity validity at high energies and is to be attributed to a dominant contribution of the  amplitude's real part. Evidently, at high energies and large values of $b$ such an assumption   faces the troubles  due to  violation of  the consequences of the  scattering matrix analyticity and unitarity and should be considered as an incorrect one. The region of  high energies includes the highest ones achieved at the LHC as well as the lower energies of CERN ISR. However, at the moment we cannot turn the qualitative estimate of the high energy region into a more quantitative one. 
  
  The  inconsistency of a peripheral dominance of the elastic scattering results, in particular, in disregard of a shadow nature of the elastic scattering at large  impact parameters. The peripheral profile dominance of $h_{el}$ masks an absence of absorption in this region.  Proper exclusion of this  option corresponds to the respective elimination of the relevant class  of the amplitude phase $t-$dependencies  and, thus, can be treated as an emergent constraints for  the phase. This exclusion, which means an elimination of the peripheral geometric elastic scattering, provides a consent with  the unitarity and analyticity and leads to a clear physical picture of the elastic scattering restoring its shadow nature at large values of the collision impact parameters.
  
 Thus,  any  phenomenological model and/or  analysis of the experimental data which exploits a nonvanishing contribution of the  amplitude real part   should control the profile of the elastic overlap function in the impact parameter representation to guarantee that  the inequality 
  \begin{equation}
  [\mbox {Re}f(s,b)]^2\ll h_{inel}(s,b)
  \end{equation}
  is valid at large values of $b$, i.e. the limiting behavior of the ratio
  \begin{equation}\label{rtin}
   h_{inel}(s,b)/\mbox {Im}f(s,b)\to 1
  \end{equation}
  takes place at large increasing  impact parameters and fixed  high  collision energy indicating
  peripherality of inelastic interactions.
  This constraint should be applied already at the model preselection stage and
 only models which obey Eq. (\ref{rtin})  can be used for the experimental data analysis. 
  
  An appearance of the constraint for the real part of the scattering amplitude is not unexpected since its real and imaginary parts are related due to dispersion relation. Full implementation of this connection is not realized at the moment, a possible  variant of such an implementation has been discussed in \cite{disp}.

  It is rather reasonable, that a  hadron  can be seen as a structure with a hard central core coated by a thick but breakable shell. Similar structure of hadrons has been proposed in \cite{islam}. As it follows from the recent studies \cite{harz, china,ji}, the mass distribution in the hadron also supports it. 
The peripheral hadron collisions  lead to the shells'
  destruction which is naturally  associated with the inelastic diffraction processes. This breaking down  the shells can therefore be a reason of the peripheral nature of the single or double inelastic diffraction processes.
  
  Diffraction on the outer shell has a shadow nature and it is observed as a diffraction peak with the first dip in 
  $d \sigma /dt$.
  Diffraction on the core has a geometrical nature  and hides the secondary dips and bumps in the differential cross--section \cite{epl}. The latter is due to the reflective scattering/hollowness which, as we believe, is being observed at the LHC. It should be noted that the latter phenomena are associated with the inelastic overlap function peripheral form at small impact parameters at the LHC energies. The central subject of this note is related to the region of large impact parameters and the dominance  of the inelastic overlap function in this region. Of course, the reflective scattering/hollowness obey this property. But the property is a more general and somehow should be observed at lower energies too, where the reflection/hollowness
  phenomenon has not been detected.
  
  The reflective scattering mode can be interpreted as a result of the central core presence in the hadron structure  responsible for appearance of the second diffraction cone in the differential cross--section of the elastic scattering \cite{epl}. This mode  implies the elastic scattering dominance at $s\to\infty$ .  Moreover, the feature of this mode is an absence of the noticeable contribution from the real part, i.e. the elastic overlap function always has a central profile and the inelastic one has a peripheral profile. 
  Presence of the central proton's core  could  be associated with a chiral symmetry spontaneous  breaking mechanism when  the two scales $\Lambda_{QCD}$ and  $\Lambda_{\chi}$  relevant for  the color confinement and spontaneous chiral symmetry breaking are different (recent discussion  is in \cite{evans}).

Finally, besides of the CNI use for the phase studies, there is  a yet another possibility of a phase-sensitive experimentation related to the polarization studies. Those are technically difficult, but  polarization measurements are  sensitive to the helicity amplitudes phases \cite{spst}. 
 \section*{Acknowledgements}
We are grateful to Laszlo Jenkovszky and Evgen Martynov for a careful reading of the note, many useful comments and remarks.
 
\section*{Figure captions} 
Figure 1: Qualitative $b$--dependencies of the elastic and inelastic overlap functions $h_{el}$  and  $h_{inel}$.\\[2ex]
Figure 2: Proton scattering  at the  impact parameter $b$.\\
============================================================
No color required.


\small
\begin{thebibliography}{99}
\bibitem{totem}
G. Antchev et al. (The TOTEM Collaboration), Eur. Phys. J. C {\bf 26}, (2016) 661.
\\
https://doi.org/10.1140/epjc/s10052-016-4399-8.
\bibitem{dre}
I.M. Dremin,  Particles {\bf 2} (2019) 57-69.
\\ https://doi.org/10.3390/particles2010005.
\bibitem{petr}
V.A. Petrov, Theor. Math. Phys. {\bf 204} (2020) 896-900.
\\ https://doi.org/10.1134/S0040577920070041.
\bibitem{khoze}
V.A. Khoze, A.D. Martin, M.G. Ryskin,  Phys. Rev. D {\bf 101} (2020) 016018. \\https://doi.org/10.1103/PhysRevD.101.016018. 
\bibitem{dur}
L. Durand, P. Ha,  Phys. Rev. D {\bf 102} (2020) 036025.
\\ https://doi.org/10.1103/PhysRevD.102.036025.
\bibitem{ferr}
  E. Ferreira, A.K. Kohara, T. Kodama, arXiv: 2011.13335v1.
\bibitem{las}
W.  Broniowski et. al. Phys. Rev. D. {\bf 98}  (2018) 074012.
\\ https://doi.org/10.1103/PhysRevD.98.074012
\bibitem{vanh}
 L. Van Hove, Rev. Mod. Phys.  {\bf 36} (1964) 655-665.
 \\ https://doi.org/10.1103/RevModPhys.36.655
 \bibitem{fsch}
B. Schrempp, F. Schrempp, Nucl. Phys. B {\bf 163}, (1980) 397.
\\
https://doi.org/10.1016/0550-3213(80)90410-1.
 \bibitem{krupa}
V. Kundr{\'a}t, M. Locaj{\'i}{\v c}ek, D. Krupa,  Phys. Lett. B {\bf 544} (2002) 132-138
\\ https://doi.org/10.1016/S0370-2693(02)02481-4.	
%\bibitem{rat13}
%G. Antchev et al. (The TOTEM Collaboration), arXiv: 1712.06153v2.
%\bibitem{chew}
%G.F. Chew, S.C. Frautchi, Phys. Rev. Lett. {\bf 5}, 580 (1960).
%\bibitem{wu}
%T.T. Wu, A. Martin, S.M. Roy, V. Singh, Phys. Rev. D {\bf 84}, 025012 (2011).
%\bibitem{fagund}
%D.A. Fagundes, M.J. Menon, P.V.R.G. Silva,
%Nucl. Phys. A {\bf 946}, 194 (2016).
%\bibitem{sachr}
%C.T. Sachrajda, R. Blankenbecler, Phys. Rev. D {\bf 12}, 1754 (1975).
\bibitem{webb}
B.R. Webber, Nucl. Phys. B {\bf 87} (1975) 269.
\\ https://doi.org/10.1016/0550-3213(75)90067-X.
\bibitem{tr}
S.M. Troshin,  Phys. Lett. B {\bf 682} (2009) 40-42.
\\
https://doi.org/10.1016/j.physletb.2009.10.088.
\bibitem{mt}
A. Martin, Phys. Rev. {\bf 129} (1963)1432-1436.
\\
https://doi.org/10.1103/PhysRev.129.1432.
\bibitem{pdbc}
P.D.B. Collins,  An Introduction to Regge Theory and High-Energy Physics, 460pp, Cambridge University Press, Cambridge 1977.
\bibitem{frs}
M. Froissart, Phys. Rev. {\bf 123} (1961) 1053-1057.
\\
https://doi.org/10.1103/PhysRev.123.1053.
\bibitem{grb}
V.N. Gribov, JETP {\bf 41} (1961) 667-669.
\bibitem{cn}
R. Cahn, Z. Phys. C {\bf 15}  (1982) 253-260.
\\
https://doi.org/10.1007/BF01475009
\bibitem{kt}
V. Kundr{\'a}t, M. Locaj{\'i}{\v c}ek, Z. Phys. C {\bf 63} (1994) 619-629.
\\
https://doi.org/10.1007/BF01557628.
\bibitem{kt1}
V. Kundr{\'a}t, M. Locaj{\'i}{\v c}ek, Mod. Phys. Lett. A {\bf 11}, 2241-2250 (1996).
\\
https://doi.org/10.1142/S021773239600223X.
\bibitem{pkan}
J. Pumplin, G.L. Kane, Phys. Rev.  D {\bf 11} (1975) 1183 .
\\
https://doi.org/10.1103/PhysRevD.11.1183.
\bibitem{ll}
G. Cohen--Tannoudji, V.V. Ilyn, L.L. Jenkovszky, Lett. Nuov. Cim. {\bf 5}, 957-962 (1972).
\\
 https://doi.org/10.1007/BF02777999.
\bibitem{halz}
F. Halzen,  Model Independent Features of Diffraction. In: Speiser D., Halzen F., Weyers J. (eds) Particle Interactions at Very High Energies. NATO Advanced Study Institutes Series (Series B: Physics), vol 4. Springer, Boston, MA.  
\\ https://doi.org/10.1007/978-1-4684-8655-1-1.
\bibitem{bl}
R. Blankenbecler, M.L. Goldberger, Phys. Rev. {\bf 126} (1962) 766-786.\\
https://doi.org/10.1103/PhysRev.126.766.
\bibitem{gold}
M.L. Goldberger, K.M. Watson, Collision Theory, John Wiley \& Sons, New-York--London--Sydney, 1964.
\bibitem{hz}
R. Henzi, Nuovo Cim. A {\bf 46}, (1966) 370.
\bibitem{schr}
B. Schrempp, F. Schrempp, Nucl. Phys. B {\bf 163} (1980) 397-452.
\\
https://doi.org/10.1016/0550-3213(80)90410-1.
\bibitem{evg}
A. Alkin et al, Phys. Rev.  D {\bf 89} (2014) 091501(R).
\\
https://doi.org/10.1103/PhysRevD.89.091501.
\bibitem{ijmpa07}
S.M. Troshin, N.E. Tyurin, Int. J. Mod. Phys. A {\bf 22} (2007) 4437-4449.
\\
https://doi.org/10.1142/S0217751X0703697X.
\bibitem{alkin1}
A. Alkin et al., arXiv: 1807.06471v2.
\bibitem{tsrg}
T.  Cs\"{o}rg\H{o}, R. Pasechnik, A. Ster,  Acta Phys. Pol. B Proc. Suppl. {\bf 12 } (2019)  779-785.
\bibitem{ro}
G. Antchev et al. (The TOTEM Collaboration),
Eur. Phys. J. C {\bf 79} (2019) 785.
\\
https://doi.org/10.1140/epjc/s10052-019-7223-4.
\bibitem{drem}
I.M. Dremin, V.A. Nechitailo, S.N. White, Eur. Phys. J. C {\bf 77} (2017) 910 .
\\
https://doi.org/10.1140/epjc/s10052-017-5483-4.
%\bibitem{asd}
%S.M. Troshin, N.E. Tyurin, Phys. Lett. B {\bf 316}, 175 (1993).
\bibitem{sam}
 V.A. Petrov, A.P. Samokhin, Int. J. Mod. Phys.: Conf. Ser. {\bf 47} (2018) 1860097.
 \\
 https://doi.org/10.1142/S2010194518600972.
 \bibitem{khr}
 V.F. Edneral et al., Preprint CERN-TH-2126, 1976.
\bibitem{plb}
 S.M. Troshin, N.E. Tyurin, Phys. Lett. B {\bf 316} (1993) 175.
 \\
 https://doi.org/10.1016/0370-2693(93)90675-8.
 \bibitem{str}
 P. Desgrolard, L.L. Jenkovszky, B.V. Struminsky, Phys. Atom. Nucl. {\bf 63} (2000) 891.
 \\
 https://doi.org/10.1134/1.855720.
 \bibitem{hdre}
 I.M. Dremin, Phys.-Usp. {\bf 58} (2015) 61.
 \\
 https://doi.org/10.3367/UFNe.0185.201501d.0065
 \bibitem{anis}
 V.V. Anisovich, V.A. Nikonov, J. Nyiri,
  Phys. Rev. D {\bf 90} (2014)  074005.
  \\
 https://doi.org/10.1103/PhysRevD.90.074005.
 \bibitem{arr}
 E.R. Arriola, W. Broniowski, Few Body Syst. {\bf 57} (2016) 485.
 \\
 https://doi.org/10.1007/s00601-016-1087-z.
\bibitem{wf}
K. Fukushima, T. Kojo, W. Weise,  Phys. Rev. D {\bf 102} (2020) 096017.
\\
https://doi.org/10.1103/PhysRevD.102.096017.
\bibitem{jpg19}
S.M. Troshin, N.E. Tyurin, J. Phys. G,  {\bf 46} (2019) 105009. 
\\
https://doi.org/10.1088/1361-6471/ab0ed1.
%\bibitem{lerr}
%L. McLerran, R.D. Pisarski, Nucl. Phys. A {\bf 796} (2007) 83-100.\\
%https://doi.org/10.1016/j.nuclphysa.2007.08.013
%\bibitem{lerr1}
%L. McLerran, Acta Phys. Pol. B {\bf 51} (2020) 1067-1977.\\
%https://doi.org/10.5506/APhysPolB.51.1067.

%\bibitem{baker}
%M. Baker, R. Blankenbecler, Phys. Rev. {\bf 128}, 415 (1962).
%\bibitem{lsl}
%W. Broniowski et al. Phys. Rev. D {\bf 98}, 074012 (2018).
%\bibitem{bron1}
%E. Ruiz Arriola, W. Broniowski, Phys. Rev. D {\bf{95}}, 074030 (2017).
%\bibitem{bron}
%W. Broniowski, E. Ruiz Arriola, Acta Phys. Polon. B {\bf{10}}, 1203 (2017).
%\bibitem{arri}
%E. Ruiz Arriola, W. Broniowski, Phys. Rev. D {\bf{95}}, 074030 (2017).
%\bibitem{soto}
%J. L. Albacete, A. Soto-Ontoso, Phys. Lett. B {\bf 770}, 149 (2017).
%\bibitem{soto1}
%J.L. Albacete, H. Petersen, A. Soto-Ontoso,  Phys.Rev. C {\bf 95}, 064909 (2017).
%\bibitem{usprd}
%S.M. Troshin, N.E. Tyurin, Phys. Rev. D {\bf 88}, 077502 (2013).
%\bibitem{khoze}
%V.A. Khoze, A.D. Martin, M.G. Ryskin, Phys. Lett. B
%{\bf 780}, 352 (2018).
%\bibitem{gsp}
%S.M. Troshin, N.E. Tyurin, Mod. Phys. Lett. A {\bf 23 }, 169 (2008).

\bibitem{burk}
V.D. Burkert, L. Elouadrhiri, F.X. Girod, Nature {\bf 557} (2018) 396-399.
\\
https://doi.org/10.1038/s41586-018-0060-z.

%\bibitem{dstot}
%G. Antchev et al. (TOTEM Collaboration) Eur. Phys. J C {\bf 79}, 861 (2019).
%\bibitem{goeke}
%K. Goeke et al. Phys. Rev. D {\bf 75}, 094021 (2007).
%\bibitem{waka}
%M. Wakamatsu, Phys. Rev. D {\bf 46}, 3762 (1992).
%\bibitem{islam}
%M.M. Islam, Z. Phys. C {\bf 53}, 253 (1992).
%\bibitem{jenk}
%L.L. Jenkovszky et al., Int. J. Mod. Phys. A {\bf 25}, 5667 (2010).
%\bibitem{gls}	
%S.V. Goloskokov, S.P. Kuleshov, O.V. Selyugin, Sov. J. Nucl. Phys. {\bf 35}, %895 (1982).
%\bibitem{chi}
%S.M. Troshin, N.E. Tyurin, Phys. Rev. D {\bf 49}, 4427 (1994).
%\bibitem{pump}
%J. Pumplin, G.L. Kane, Phys. Rev. D {\bf 11}, 1183 (1975).
%\bibitem{royzen}
%I.I. Royzen, E.L. Feinberg, O.D. Chernavskaya, Phys. Uspekhi {\bf {47}}, 427 %(2004).
%\bibitem{petr}
%R. Petronzio, S. Simula, G. Ricco, Phys. Rev. D {\bf 67}, 094004 (2003).
%\bibitem{gold}
%T. Goldman, R.W. Haymaker, Phys. Rev. D {\bf 24}, 724 (1981).
%\bibitem{low}
%F.E. Low, Phys. Rev. D {\bf 12}, 163 (1975). 
%\bibitem{orear}
%J. Orear, Phys. Rev. D {\bf 18 }, 2484 (1978).
%\bibitem{heines}
%G.W. Heines, M.M. Islam, Nuov. Cim. {\bf 61}, 149 (1981).
%\bibitem{chern}
%I.M. Dremin, D.S. Chernavsky, Sov. Phys. JETP {\bf 11 }, 167 (1960).
%\bibitem{ech}
%S.M. Troshin, N.E. Tyurin, Fiz. Elem. Chast. Atom. Yadra {\bf 15}, 53 (1984).
%\bibitem{ans}
%A.A. Anselm, I.T. Dyatlov,
%Phys. Lett.  B {\bf 24}, 479 (1967).
%\bibitem{anddr}
%I.V. Andreev, I.M. Dremin, JETP Lett. {\bf 6}, 262 (1967).
%\bibitem{sav}
%V.I. Savrin, N.E. Tyurin, O.A. Khrustalev, Yad .Fiz. {\bf 10}, 856 (1969).
%\bibitem{7tev}
%S.M. Troshin, N.E. Tyurin, Mod. Phys. Lett. A {\bf 27}, 1250111 (2012).
%\bibitem{bras}
%V.P. Gon{\c c}alves, P.V.R.G. Silva,  Eur. Phys. J. C {\bf 79}, 327 (2019).
%\bibitem{tsrg1}
%T.  Cs\"{o}rg\H{o} et al.,   arXiv:1912.11968.
%\bibitem{dremn}
%I.M. Dremin, MDPI Particles {\bf 2}, 57 (2019).
%\bibitem{geom}
%S.M. Troshin, N.E. Tyurin, Phys. Rev. D {\bf 88}, 077503 (2013).
%\bibitem{19}
%S.M. Troshin, N.E. Tyurin, Mod. Phys. Lett. A {\bf 33}, 1950259 (2019).
\bibitem{del}
S.M. Troshin, N.E. Tyurin, Phys. Lett.  B {\bf 707}  (2012) 558-561.
\\
https://doi.org/10.1016/j.physletb.2012.01.033.
\bibitem{prr}
S.M. Troshin, N.E. Tyurin, Phys. Rev.  D {\bf 88},  077502 (2013).
\\
https://doi.org/10.1103/PhysRevD.88.077502.
\bibitem{epl}
S.M. Troshin, N.E. Tyurin, EPL {\bf 129}, 31002 (2020).
\\
https://doi.org/10.1209/0295-5075/129/31002.

%\bibitem{gol}
%S.V. Goloskokov, S.P. Kuleshov, O.V. Selyugin, Z. Phys. C {\bf 50}, 455 (1991).
%\bibitem{isl}
%M.M. Islam,  Z. Phys. C {\bf 53}, 92 (1991) 253.
%\bibitem{trt}
%S.M. Troshin, N.E. Tyurin, Phys. Rev. D {\bf 49}, 4427 (1994).
\bibitem{disp}
S.M. Troshin, N.E. Tyurin, Mod. Phys. Lett. A {\bf 32} (2017) 1750028 .
\\
https://doi.org/10.1142/S0217732317500286.
%\bibitem{samo}
%V.A. Petrov, A.P. Samokhin,  Int. J. Mod. Phys. Conf. Ser. {\bf 47},  1860097 %(2018).
\bibitem{islam}
M.M. Islam, Nucl. Phys. B (Proc. Suppl.) {\bf 25} (1992) 104. 
\bibitem{harz}
D.E. Kharzeev, arXiv:2102.00110v2.
\bibitem{china}
R. Wang, W. Kou, X. Chen, arXiv:2102.01610v2.
\bibitem{ji}
X.-D. Ji, arXiv:2102.07830v1.
\bibitem{evans}
N. Evans, K.S. Rigatos,  arXiv: 2012.00032v1.
\bibitem{spst}
M.G. Echevarria et al.,  PoS SPIN2018 (2019) 063.
\\
https://doi.org/10.22323/1.346.0063.







%\bibitem{difs}
%F. Ravera (The TOTEM Collaboration), 134th LHCC meeting, 30 May 2018; 
%\bibitem{difs1}
%F. Nemes  (The TOTEM Collaboration), 4th Elba Workshop on Forward Physics@LHC energy, 24-26 May 2018.
%\bibitem{pl}
%S.M. Troshin, Phys. Lett. B {\bf  682},  40  (2009).
%\bibitem{nic}
%L. Lukaszuk, B. Nicolescu, Nuov. Cim. Lett. {\bf 8}, 405 (1973).



%\bibitem{khuri}
%N.N. Khuri, in Proc. of Vth Blois Workshop---Int. Conf. Elastic and Diffractive Scattering, Providence, RI, June %8-12, 1993, eds. H.M. Fried, K. Kang, and C.-I. Tan (World Scientific, 1994), p. 42. 
%\bibitem{drem}
%I.M. Dremin, V.A. Nechitailo,
%Eur. Phys. J. C {\bf78}, 913 (2018).
%\bibitem{white}
%I.M. Dremin, V.A. Nechitailo, S.N. White, 
%Eur. Phys. J. C {\bf 77}, 910 (2017).
%\bibitem{dremin}
%I. M. Dremin, Phys. Uspekhi {\bf {58}}, 61 (2015).

%\bibitem{srvalue}
%S.M. Troshin, N.E. Tyurin, Phys. Lett. B {\bf  316},  175 (1993).
%\bibitem{schr}
%B. Schrempp, F. Schrempp, Nucl. Phys. B {\bf 163}, 397 (1980). 

%\bibitem{blr}
%S.M. Troshin, N.E. Tyurin, Mod. Phys. Lett. A {\bf 31}, 1650025 (2016).




%\bibitem{meis}
%U.G. Meissner, Phys. Rep. {\bf 161}, 213 (1988).
%\bibitem{ore}
%J. Orear, Phys. Rev. D {\bf 18}, 2484 (1978).
%\bibitem{krisch}
%A.D. Krisch, Phys. Rev. Lett. {\bf 11}, 217 (1963).







%\bibitem{cent}
%S.M. Troshin, N.E. Tyurin,  
%Int. J. Mod. Phys. A {\bf34}, 1950172  (2019). 
%\bibitem{asp}
%S.M. Troshin, N.E. Tyurin,  Mod. Phys. Lett. A {\bf 33}, 1850040 (2018).
%\bibitem{ble}
%M. Bleszynski, R.J. Glauber, P. Osland, Phys. Lett. B {\bf 104}, 389 (1981).
%\bibitem{and}
%A.F. Andreev, Sov. Phys. JETP {\bf 19}, 1228 (1964).
%\bibitem{sadz}
%M. Sadzikowski, M. Tachibana,  Acta Phys. Polon. B {\bf{33}}, 4141 (2002). 
%\bibitem{wu}
%R. Gastmans, S.L. Wu, T.T. Wu, Phys. Lett. B {\bf 720}, 205 (2013). 
%\bibitem{ch}
%G.F. Chew, S.C. Frautchi, Phys. Rev. Lett. {\bf 5}, 580 (1960).
\newpage
\end{thebibliography}
\end{document}